\documentclass[preprint]{revtex4-2}
\usepackage[english]{babel} % English language/hyphenation
\usepackage{amsmath,amsfonts,amsthm,bm} % Math packages
\usepackage{graphicx,epsfig,subfigure}
\usepackage{gensymb}
\usepackage{dcolumn}% Align table columns on decimal point
\usepackage{bm}% bold math
\usepackage{color}
\usepackage[autostyle]{csquotes}
\usepackage{tikz}
\usepackage{setspace}
\usepackage{multirow}
\usepackage{hyperref}
\usepackage{dsfont}
\usepackage{upgreek}
\usepackage{lineno}

\usepackage[latin1]{inputenc}

\begin{document}
%\preprint{APS/123-QED}
%\preprint{}

\title{Velocity jump process with volume exclusions in a narrow channel}


\author{Gayani Tennakoon}
\email[]{gten341@aucklanduni.ac.nz}
%\thanks{}
\author{Stephen W. Taylor}
\email[]{s.taylor@auckland.ac.nz}
\affiliation{Department of Mathematics, University of Auckland, Auckland 1010, New Zealand.}


\date{\today}

\begin{abstract}
This paper analyses the impact of collisions in a system of $N$ identical hard-core particles driven according to a velocity jump process. The physical space is essentially a channel in $\mathds{R}$ with a probability of occupants being able to pass each other. The system mimics what nature does, where individuals pass one another in a narrow channel while making incidental contact with those moving in the opposite direction. The passing probability may depend on the particles' size and the channel's width. Starting from the particle level model, we systematically derive a nonlinear transport equation based on an asymptotic expansion. Under low-occupied fractions, numerical solutions of both the kinetic model and the stochastic particle system are compared well during biased and unbiased random velocity changes. Analysis of the subpopulation motility within a large population exhibits the consequences of volume exclusions and channel confinements on the travelling speeds.   
\end{abstract}


\maketitle

\newpage
\raggedbottom
\section{Introduction}
Many physical and biological systems consist of individuals with collective behaviour under confined conditions. Examples where particles are driven in a domain confined to a narrow channel, include molecular and ion transport through bacterial porins \cite{nestorovich2002designed} or nuclear pore complexes [\citealp{bednenko2003nucleocytoplasmic}, \citealp{rout2003virtual}], the floating seeds spreading in vegetated open channels \cite{defina2010floating} and polymer solutions [\citealp{gutsche2008colloids},\citealp{kruger2009diffusion}]. The assumptions in a narrow channel are reasonable to model even pedestrian motion [\citealp{pellegrini2010wrong},\citealp{piccoli2011time}]. In all these applications an additional factor comes into play if the occupying individuals have a finite size or at least tend to keep others at a distance. Understanding the interplay between these constraints and particle motion is essential to explain the systems' global behaviour. \\
A classical approach for modelling random dispersal due to sudden changes in velocity, such as in a bacterial population, animal or robot swarm, is a velocity jump process  \cite{othmer1988models}. In a velocity jump process, stochastic changes are applied to the velocity rather than the position. A particle may change its current velocity $\textbf{v} \in V \subset \mathds{R}^d$ at a small time-step $dt$ with probability $\lambda dt$, where $\lambda$ is the turning frequency. Given that a jump occurred, a turning kernel $T(\textbf{v},\textbf{u})$ defines the probability of a change in velocity from $\textbf{u}$ to $\textbf{v}$ (in the same velocity space $V$).
In light of the above information, the evolution of the density function $p(\textbf{x},\textbf{v},t)$ for individuals in 2d-dimensional ($d=1,2,...$) phase space with coordinates $(\textbf{x},\textbf{v})$ is governed by the transport equation  
\begin{align} \label{VelocityJump}
\begin{split}
\frac{\partial p}{\partial t} + \nabla_{\textbf{x}}\cdot\textbf{v}p(\textbf{x},\textbf{v},t) = \lambda\int\limits_{V}T(\textbf{v},\textbf{u})p(\textbf{x},\textbf{u},t)d\textbf{u} \\ - \lambda p(\textbf{x},\textbf{v},t), \quad t\geq 0
\end{split}
\end{align} 
where $\textbf{x} \in \Omega \subset \mathds{R}^d$. This model best describes the motion of flagellated bacteria such as E. coli that possess two behavioural modes; runs and tumbles; nevertheless, there are applications to locust nymphs march \cite{erban2011individual}, bird movement \cite{taylor2015birds} and robotic systems \cite{taylor2015mathematical}. In one space dimension, when particles possess random motions with a constant speed and switch directions at an instantaneous time with an unbiased constant reversal rate, this model recovers one of the earliest correlated random walk models proposed by Goldstein and Kac [\citealp{goldstein1951diffusion}, \citealp{kac1974stochastic}]. The limitation is that many of the earliest velocity-jump processes are noninteracting; hence crowding effects are not considered. This means overlaps are permitted, which is unreal in a dense population. More recently, though, there has been an increasing interest in understanding crowding effects when accounting for the finite size of particles [\citealp{franz2016hard}, \citealp{ralph2020one}]. \\
The impenetrable finite-sized hard-core particles give rise to the so-called excluded-volume effect. Unlike non-interacting point particles, this exclusion reduces the free space in a system and influences the transport properties of diffusing particles, especially in crowded environments. In the extreme case in which the particles' diameter is equal to the width of the channel, bypassing is forbidden \cite{macey1967time}. When introducing a theoretically justified framework to analyse hard-core systems, one must choose the most appropriate particle representation with interactions. A common approach has been incorporating volume exclusion to lattice-based random walk models restricting particles' motion to a grid. During on-lattice interactions, the target site is occupied by, at most, a single particle and evolves according to a set of rules based on the neighbouring sites' condition. The simplest model is a random walk on the one-dimensional lattice with jumps (left/right) to the nearest vacant site \cite{treloar2011velocity}. However, these lattice-based models may not be realistic in some scenarios, as the mechanism allows individuals to jump across. Also, the approximation restricts the choice of the model parameters and the initial condition. Alternatively, one can consider a more realistic lattice-free random walk, in which the individual changes its position in a continuous space rather than restricting it to a lattice. The volume exclusion can be introduced by assuming particles are hard spheres that cannot overlap \cite{bruna2014diffusion}, an attempt-and-abort mechanism based on a moving probability \cite{irons2016lattice} or denying the cells to pass an equilibrium distance \cite{murray2012classifying}.  \\
Once the particles' physical representation is determined, the next concern would be the model's description level, whether particle-level or population-level. The former treats each occupant as a discrete entity and describes their behaviour explicitly. This behaviour may include internal processes and interactions between individuals, which are mathematically explained by an evolution update rule given in the form of an algorithm or differential equation. Discrete particle-level models may be conceptually simple but demand expensive computer simulations when administering them to large particle systems with complex behavioural patterns. On the other hand, continuum population-level models are responsive to large numbers and relatively easy to analyse and solve. Since they consider group-level quantities rather than individual properties, they might not capture details at particle level \cite{ryan2016model}. Continuum population-level models most commonly use partial differential equations (PDEs). For the derivation, one must consider the system variables such as number density or spatial population density. Connecting the two levels of descriptions is challenging and not evident in general, especially when the system includes particle-particle and particle-environmental interactions. Indeed one can either use a particle-level or population-level model depending on the subject of relevance and available experimental data. Nevertheless, it is essential to understand the link to get an insight into the dispersing systems.\\
Reviewing previous studies on collective dynamics and self-organisation within narrow channels, we find many mathematical and computational efforts focused on non-interacting velocity jump processes \cite{zhou2021distribution}, stochastic models [\citealp{defina2010floating},\citealp{humenyuk2020separation}] or continuum population-level models [\citealp{bruna2014diffusion},\citealp{ai2013transport}] developed for Brownian colloidal. To this end, we propose and examine an interacting velocity jump process confined to a narrow channel. The approach is to tackle the common first steps in any such problems rather than focusing on a particular question linked to studies presented earlier. The system requires advanced mathematical frameworks to capture fundamental characteristics; hence, this paper aims to determine the population-level model systematically via the particle-level description. We develop a method similar to the study \cite{ralph2020one} and analyse the impact of collisions between individuals on the behaviour of groups of particles, driving according to a velocity jump process. In particular, we too examine a hard-core $N$-particle system, though we consider a domain confined to a narrow channel and wide enough for particles to pass each other. The occupants exclude a volume in the channel, but in the passing regime, the region is some sort of an interface that is no longer excluded. We still have collisions with the particles moving in opposite directions that are successful with probability given by a function of the particle's size and the width of the channel. The collision probability turns up in the nonlinear transport term of the resulting hyperbolic equations and extends the work accomplished in \cite{ralph2020one}. \\
The paper is structured as follows: we begin in section (\ref{sec_IBM}) by writing down the particle-based description of the system, which consists of an N-dimensional PDE for the joint probability density function in the probability space. The following section (\ref{sec_popu}) then derives equations for the population-level behaviour that arise from those particle-level dynamics. Sections (\ref{sec_time}) and (\ref{sec_stat}) are devoted to investigate the validity of the derived model by comparing its solutions with stochastic simulations of the full particle system. Here we present several numerical examples to demonstrate the behaviour of both point and finite-size particles under different external influences.      

\section{Particle-level description}\label{sec_IBM}
\begin{figure}
\centering 
\subfigure[]{\label{single_file}\includegraphics[scale=0.8]{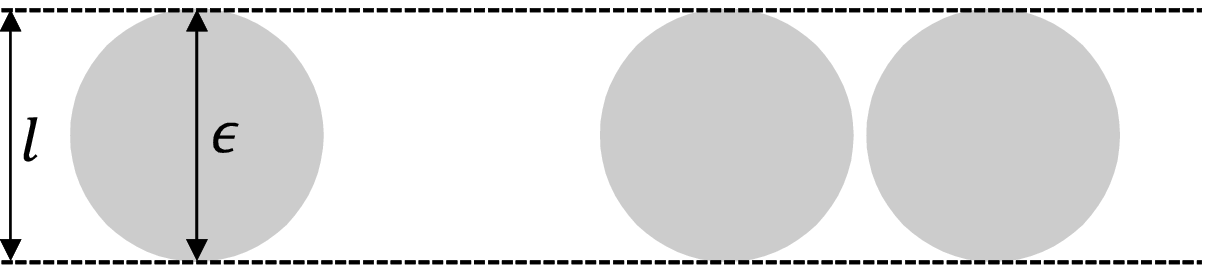}}
\subfigure[]{\label{narrow_channel}\includegraphics[scale=0.8]{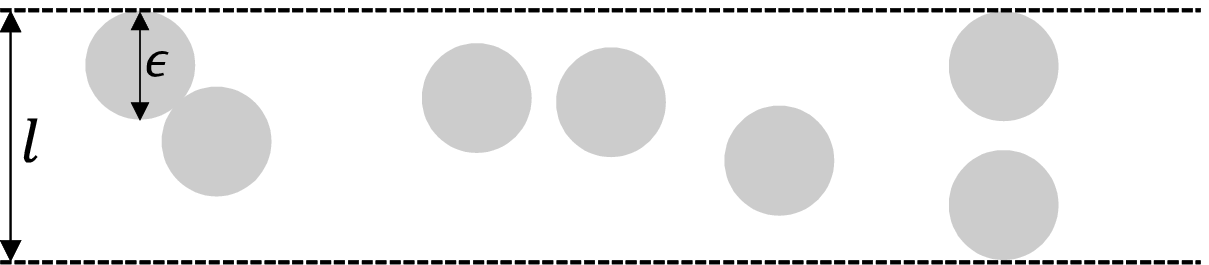}}
\caption{(a) A single-file channel where particle's diameter is the same as the width of the channel. (b) A narrow channel where particles crossover and easily change order.} {\label{single_file_narrow}} 
\end{figure}

In a previous paper, Ralph \textit{et al.} \cite{ralph2020one} analysed the velocity jump process in a single-file channel in which particles preserve the initial order over time (see Fig.\ref{single_file}). To make this paper self-contained, we summarise their particle-level description in the nondimensional form and study the collective behaviour in the passing regime (see Fig.\ref{narrow_channel}). For ease of reference, let us name the former a \textit{collision system} and the latter a \textit{narrow channel system}. \\
We, too, consider a stochastic system in compact velocity space with constant speed. We neglect the effect of background noise and assume that the system does not have external forces which restrict or prevent the particles' motion and turn velocity back to zero. The system has $N$ interacting particles in a one-dimensional domain of length $L$ that does not change through time, meaning there is no birth/death. The particles are identical hard disks (or spheres) of diameter $\epsilon$ instead of being hard rods. The stochastic changes are applied to each particle's velocity rather than to its position in space; therefore, a particle switches its direction based on $N$ independent Poisson processes with rates $\lambda(x,v)>0$.
Collisions with another particle or with the domain wall may also change the velocity; however, we expect fewer collisions when the domain dimensions are larger than the particle's size. Furthermore, the wall-particle interactions are limited to the domain's left and right ends due to their unidirectional motion. \\
The equivalent PDE description in terms of the joint probability density function $P(\vec{x},\vec{v},t)$ for $N$ particles to be found at the position $\vec{x}=(x_1,...,x_N)$ for $x_i\in\Omega(=[0,L])$ with velocity $\vec{v}=(v_1,...,v_N)$ for $v_i\in V(=\{-c,c\})$ at time $t$, given by
\begin{align}\label{IBTP1}
\begin{split}
\frac{\partial P}{\partial t} + \vec{v}\cdot\nabla_{\vec{x}}P + \sum_{i=1}^{N}\bigl(\lambda(x_{i},v_{i})P(\vec{x},\vec{v},t) - \\ \lambda(x_{i},-v_{i})P(\vec{x},s_{i}\vec{v},t)\bigr) = 0
\end{split}
\end{align}
as in \cite{ralph2020one}. The operator $s_i$ switches velocity of the $i^{th}$ particle and the random turning rate $\lambda(x_i,v_i) = \lambda_0 - \chi v_iD_{x_i}S(x_i)$ expresses the knowledge on how correlation depends on an extracellular signal $S$ (attractant or repellent) when the sensitivity coefficient is $\chi$. In the passing regime, the excluded area $\{\vec{x}\in\Omega^N: |x_i-x_j|\leq\epsilon, \forall i\neq j\}$ act as an interface where $P$ has jumps at $x_i\pm\epsilon$, $\forall i$. We suppose all the hard-core particles are distributed independently and identically in the domain initially; however, they may not preserve their initial ordering. Inside the domain, particles collide with probability $\delta$, otherwise move independently. This collision probability is another small parameter that depends on the particle size and the channel width (say $l$) in a particular modelling situation. Here, we treat $\delta$ as an independent parameter. During a bypass, a particle has access to another's excluded region from the left and right ends; consequently, the collision boundary condition of \cite{ralph2020one} is replaced by an interface condition. We detail these conditions in the following section. 

\section{Population-level model} \label{sec_popu}
The aim is to reduce the higher-dimensional PDE for the joint density $P(\vec{x},\vec{v},t)$ to a low-dimensional PDE for the marginal density $p(x,v,t)$ of a single particle. This dimension reduction can be executed in the same manner as in \cite{ralph2020one}, except now we evaluate $P$ using the interface conditions. The simplest case is when $\epsilon=0$, where the domain $\Omega^N$ has no holes and particles are independent. This yields an equation
\begin{subequations} \label{PP}
\begin{align}
\begin{split}
\frac{\partial{p}}{\partial{t}} +v\frac{\partial{p}}{\partial{x}} + &\lambda(x,v)p(x,v,t)- \\ &\lambda(x,-v)p(x,-v,t) = 0 \quad\text{in}\quad\Omega 
\end{split}\\
p(x,v,t) &= p(x,-v,t)\quad\text{on}\quad\partial\Omega
\end{align} 
\end{subequations}
that does not contain any confinement parameters. In fact, it is similar to that of Goldstein and Kac [\citealp{goldstein1951diffusion}, \citealp{kac1974stochastic}], except now the turning rate is a variable. \\
When $\epsilon>0$ a particle excludes a space leading to a domain $\Omega^N$ with holes; the individuals become no longer independent. We find configurations in which two or more particles are approaching each other; however, in the low volume fraction regime, the volume in the integration space occupied by two particles dominates \cite{bruna2012excluded}. Therefore, we fix two particles at $x_1$ with $v_1$ and $x_2$ with $v_2$ to illustrate our approach. It is important to note that the immediate difference from a collision system configuration space is that $x_2 \in (x_1-\epsilon,x_1+\epsilon)$ is  not a constant illegal configuration anymore. That is, particle $2$ passes the inner region of particle $1$ with probability $1-\delta$; otherwise, bounces back. Hence, the space available for particle $2$ centre is still $\Omega(x_1)=[0,L]$ but with discontinuities in its density function at $x_2 = x_1\pm\epsilon $. Taking this into account, we write the following interface conditions near interfaces: 
\begin{subequations} \label{IC} 
\begin{align} \label{IC_left}
\text{when $x_2 < x_1$}\\
\begin{split}
P(x_1,x_1^--\epsilon,c,-c,t)&=\delta P(x_1,x_1^--\epsilon,-c,c,t)+P(x_1,x_1^+-\epsilon,c,-c,t)  \\
P(x_1,x_1^+-\epsilon,-c,c,t)&=(1-\delta)P(x_1,x_1^--\epsilon,-c,c,t) 
\end{split}
\end{align}
\begin{align} \label{IC_right}
\text{when $x_2 > x_1$}\\
\begin{split}
P(x_1,x_1^++\epsilon,-c,c,t)&=\delta P(x_1,x_1^++\epsilon,c,-c,t)+P(x_1,x_1^-+\epsilon,-c,c,t) \\
P(x_1,x_1^-+\epsilon,c,-c,t)&=(1-\delta)P(x_1,x_1^++\epsilon,c,-c,t)
\end{split}
\end{align}
\end{subequations}
where $\delta\equiv\delta(\epsilon,l)$. To derive the equation for the marginal density $p$, we now integrate the two-particle form of equation (\ref{IBTP1}) over $\Omega(x_1)\times V$, rewriting the domain over subintervals $[0,x_1-\epsilon)$, $(x_1-\epsilon,x_1+\epsilon)$ and $(x_1+\epsilon,L]$ where necessary. This integration result in equation
\begin{equation}
\begin{aligned} \label{IBTP2}
\frac{\partial{p}}{\partial{t}}+v_1\frac{\partial{p}}{\partial{x_1}}+ 2v_1\left( P(x_1,x_2,v_1,-v_1,t)|_{x_2=x_1^-+\epsilon}^{x_2=x_1^++\epsilon}+P(x_1,x_2,v_1,-v_1,t)|_{x_2=x_1^--\epsilon}^{x_2=x_1^+-\epsilon}\right) &+ \\\lambda(x_1,v_1)p(x_1,v_1,t) - \lambda(x_1,-v_1)p(x_1,-v_1,t) &= 0 
\end{aligned}
\end{equation} 
involving the two-particle density which is confined to the interaction interfaces. The evaluation of this unknown term requires a rational approach instead of any ad hoc closure approximations. Here we use an asymptotic expansion that allows us to find an approximate $P$ which does not breaks down in the interval $(x_1-\epsilon,x_1+\epsilon)$. The outer solution $P(x_1,x_2,v_1,v_2)$ can be simply define as 
\begin{subequations} \label{Outer}
\begin{align}
\begin{split}
P_l(x_1,x_2,v_1,v_2,t) = q(x_1,v_1,t)q(x_2,v_2,t) &+ \epsilon P_l^{(1)}(x_1,x_2,v_1,v_2,t) + ..., \\ &\text{for}\quad{x_2\in}[ 0, x_1-\epsilon )
\end{split}\\
\begin{split}
P_r(x_1,x_2,v_1,v_2,t) = q(x_1,v_1,t)q(x_2,v_2,t) &+ \epsilon P_r^{(1)}(x_1,x_2,v_1,v_2,t) + ..., \\ &\text{for}\quad{x_2\in}( x_1+\epsilon, L ]
\end{split}
\end{align}
\end{subequations}
for some distribution function $q$. The normalization condition on $P$ estimates $q({x}_1,v_1,t)=p({x}_1,v_1,t)+\mathcal{O}(\epsilon)$. In the correlated region $(x_1-\epsilon,x_1+\epsilon)$, we may define
$P\equiv P_{in} $ as a series in the small parameter $\epsilon$. This is because the leading order term in the outer solutions (\ref{Outer}) and inner solution $P_{in}$ satisfies the interface conditions (\ref{IC}). Moreover, using the fact that particles are identical and indistinguishable, we find   
\begin{equation}
\begin{split}
P(x_1,x_2,v_1,-v_1,t)|_{x_2=x_1^-+\epsilon}^{x_2=x_1^++\epsilon}+P(x_1,x_2,v_1,-v_1,t)|_{x_2=x_1^--\epsilon}^{x_2=x_1^+-\epsilon} = \\ \delta\epsilon\frac{\partial}{\partial{x_1}}\left[q({x}_1,v_1,t)q({x}_1,-v_1,t)\right]
\end{split}
\end{equation}
Recall that we began our analysis considering pairwise interactions at $\mathcal{O}(\epsilon$) which can
now be easily extended to $N$ particles. Hence, the density $p(x,v,t)$ satisfies the kinetic equation
\begin{align} \label{kinetic_1}
\begin{split}
\frac{\partial{p}}{\partial{t}}+v\frac{\partial{p}}{\partial{x}}+ 2v\delta\epsilon(N-1)\frac{{\partial}}{\partial x}{p}{p(x,-v,t)} +\\ \lambda(x,v)p(x,v,t)-\lambda(x,-v)p(x,-v,t) = 0
\end{split}
\end{align}
with the boundary condition 
\begin{equation} \label{BC}
p(x,v,t) = p(x,-v,t)\quad\text{at}\quad x=0,L
\end{equation}
The nonlinear PDE is fairly similar to the model in previous work \cite{ralph2020one}. The additional term that depicted the effect of pairwise interactions in the full collision system now incorporates a collision probability. Although we assumed it as small, it may take any value between zero and one as far as the model concerns; the maximum value resembles the collision system (the derivation of $\delta$ for a particular $\epsilon$ and $l$ is given in Appendix \ref{Collision_Prob}). A higher number of collisions, resulting from a larger $N$ or $\epsilon$ may obstruct the motion toward favourable directions. One can think of this model as an off-lattice version examined in \cite{treloar2011velocity}; however, the reduced continuum model we have obtained for the population-level behaviour differs from the corresponding continuum limit of the discrete on-lattice counterpart model. Specifically, the nonlinear transport terms of the coupled system of hyperbolic PDEs obtained due to crowding effects do not agree with those derived in our model. \\
To assess the validity of the model (\ref{kinetic_1}), we compare the numerical results of our continuum model for the narrow channel to the results from the corresponding particle-level model as well as the continuum models of point particles and collision systems. The analysis will explain the conditions under which the model can describe population-level behaviour emerging from the particle-level dynamics.

\section{Time-dependent solutions} \label{sec_time}
We study solution strategies based on characteristics for time-dependent hyperbolic balance laws. Rather than adhering to standard numerical methods, this approach is comprehensive and practical. Numerous work has contributed to understanding nonlinear hyperbolic systems of equations, such as the shallow water equations \cite{abdelrahman2017shallow} and compressible Euler equations \cite{osher1983upwind}. The main ingredients in the study of such systems are the concepts of characteristics and Riemann invariants. We apply them for both the theoretical and computational developments in the system we study. \\
By rewriting the equation (\ref{kinetic_1}) for subpopulation densities $p^+=p(x,c,t)$ and $p^-=p(x,-c,t)$ yield the following hyperbolic system:
\begin{equation} \label{conservative_form}
\frac{\partial\vec{p}}{\partial t}+\frac{\partial}{\partial x} F(\vec{p}) = \vec{g}(x,\vec{p}), \quad 0\leq x \leq L, \quad 0\leq t \leq T
\end{equation}
where $\vec{p} = \begin{pmatrix} p^+\\ p^- \end{pmatrix}$, $ F(\vec{p}) = \begin{pmatrix} cp^++c\xi p^+p^- \\ -cp^--c\xi p^-p^+ \end{pmatrix} $ with $\xi = 2\delta\epsilon(N-1)$ and the source term $\vec{g}(x,\vec{p}) = \begin{pmatrix} \lambda^-(x)p^- - \lambda^+(x)p^+ \\ \lambda^+(x)p^+ - \lambda^-(x)p^- \end{pmatrix}$. When $\epsilon=0$, the Jacobian produces linearly degenerate fields 
\begin{align} \label{eig_charac_point}
\left\{\left[c,\left(\begin{array}{c} 1\\ 0\end{array}\right)\right],\left[-c,\left(\begin{array}{c} 0\\ 1\end{array}\right)\right]\right\}
\end{align}
that reduces the noninteracting system (\ref{PP}) into a simple system of first-order ODEs:
\begin{align} \label{ODE_point}
\begin{split}
\frac{dp}{dt}^+ &= g_1(x,\vec{p})\quad\text{along}\quad x = ct+x_0 \\
\frac{dp}{dt}^- &= g_2(x,\vec{p})\quad\text{along}\quad x = -ct+x_0
\end{split}
\end{align}
with the initial condition $p(x,\pm c,0)$ and the reflective boundary condition $p^+=p^-$ at $x=0,L$. This system can be solved by employing a numerical integration method with a fixed time step, which is not the case in a nonlinear system. \\
When $\epsilon\neq 0$, for every $\vec{p}\in \mathds{R}^2$ we find two distinct real eigenvalues paired with two linearly independent eigenvectors. But the structure of these eigenvectors does not provide much help for the latter computations; alternatively, we consider the periodic extensions for the marginal densities: $u_1(x,t)$, the odd extension of $p^+-p^-$, and $u_2(x,t)$, the even extension of $p^++p^-$, where the boundary condition (\ref{BC}) extends $u_1$ and $u_2$ as continuous periodic functions with period $2L$. The solution domain changes to $[-L, L]$, and the non-conservative system reads as
\begin{align}\label{conserv_law}
\begin{split}
\frac{\partial\vec{u}}{\partial t}+J(\vec{u})\frac{\partial\vec{u}}{\partial x} = \vec{g}(x,\vec{u}), \quad -L\leq x \leq L \\
\text{with the periodic boundary condition}\\ \vec{u}(-L,t) = \vec{u}(L,t),
\end{split}
\end{align}
where $\vec{u} = \begin{pmatrix} u_1\\ u_2 \end{pmatrix}$, $J(\vec{u}) = \begin{pmatrix} -c\xi u_1 & c(1+\xi u_2) \\ c & 0 \end{pmatrix}$ and the source term $\vec{g}(x,\vec{u}) = \begin{pmatrix} \mu_1u_2 + \mu_2u_1 \\ 0 \end{pmatrix}$ with $\mu_1=\lambda^-- \lambda^+$ and $\mu_2=-\lambda^-- \lambda^+$. We use this extended system to analyse time-dependent solutions with and without random changes in the velocities. Essentially we expect to get two ODEs for some algebraic combinations of $u_1$ and $u_2$ along the characteristic curves. 

\subsection{No random transitions: \texorpdfstring{$\lambda^{\pm} \equiv 0$}{Lg}} 
When particles do not experience random changes in the direction, the source term vanishes; the solutions of (\ref{ODE_point}) are simply the initial distributions travelling to right and left at constant speed $c$. In an interacting system, we still find velocity changes due to collisions, but we expect the nonlinear model to behave like the point-particle linear model at lower values of collision probabilities. In fact, the solution procedure becomes more involved in the presence of nonlinear transport terms. From the system (\ref{conserv_law}), we first derive the characteristic ODEs from the eigenvalues and their associated eigenvectors. We then find functions that are invariant along the characteristic directions and satisfy a set of convective equations. \\
The eigenvalues 
\begin{align} \label{evalue1}
\begin{split}
\Lambda = c(a\pm \sqrt{b}) \quad\text{of}\quad J, \quad\text{where}\\ a=-\frac{\xi u_1}{2} \quad\text{and}\quad b = 1+\xi u_2+\frac{1}{4}(\xi u_1)^2
\end{split}
\end{align}
represent the characteristic directions at which hard particles propagate, paired with the left eigenvectors
\begin{align} \label{evec1}
V = (1, -a\pm\sqrt{b})
\end{align}
The system is strictly hyperbolic, because the eigenvalues are all real and distinct, as long as $b$ remains positive. Also, the $\Lambda_i$-characteristic field is genuinely nonlinear as 
\begin{align*}
\nabla\Lambda_1(u_1,u_2)\cdot V_1 &= \frac{c\xi^2u_1}{2\sqrt{b}}\quad\text{and}\quad\\
\nabla\Lambda_2(u_1,u_2)\cdot V_2 &= -\frac{c\xi^2u_1}{2\sqrt{b}} \quad\text{for}\quad \xi \neq 0.
\end{align*}
When the hyperbolic system is multiplied by the eigenvectors (\ref{evec1}), the right-hand side of (\ref{conserv_law}) condenses and collapses down to the following two ODEs along the characteristics $D_tx = \Lambda_i(u_1,u_2)$:
\begin{equation}  \label{ODE}
V_{i1}\frac{du_1}{dt}+V_{i2}\frac{du_2}{dt} = 0 \quad \text{for}\quad i = 1,2
\end{equation} 
Since the asymptotic expansion is accurate upto $\mathcal{O}(\epsilon)$, we expand $\sqrt{b}$ and avoid higher order terms. This approximation simplifies the above ODEs into
\begin{align*} 
&\frac{du_1}{dt}+\big(1+{\textstyle\frac{\xi}{2}}(u_1+u_2)\big)\frac{du_2}{dt} = 0, \\
&\frac{du_1}{dt}+\big({\textstyle\frac{\xi}{2}}(u_1-u_2)-1\big)\frac{du_2}{dt} = 0,
\end{align*}
which are integrable using an integrating factor. Hence, it follows that the Riemann invariants are  
\begin{align} \label{R1R2}
\begin{split}
\mathcal{R}_1(u_1,u_2) = (u_1+u_2)e^{\frac{\xi u_2}{2}} \quad\text{on characteristics}\\ x(t) = c\big(1+{\textstyle\frac{\xi}{2}}(u_2-u_1)\big)t + x_0, \\
\mathcal{R}_2(u_1,u_2) = (u_1-u_2)e^{\frac{\xi u_2}{2}} \quad\text{on characteristics}\\ x(t) = -c\big(1+{\textstyle\frac{\xi}{2}}(u_2+u_1)\big)t + x_0 
\end{split}
\end{align} 
It is possible to obtain an exact solution for $p^+$ and $p^-$ in the nonlinear system (\ref{kinetic_1}) during this unbiased situation. Since $\mathcal{R}_1$ and $\mathcal{R}_2$ are constant along their respective characteristics, given the initial conditions, say $u_1^0(x_0)$ and $u_2^0(x_0)$, we write
\begin{align*} 
\begin{split}
\mathcal{R}_1(u_1(x,t),u_2(x,t)) = \mathcal{R}_1(u_1^0(x - \Lambda_1t),u_2^0(x - \Lambda_1t)) \\
\mathcal{R}_2(u_1(x,t),u_2(x,t)) = \mathcal{R}_2(u_1^0(x - \Lambda_2t),u_2^0(x - \Lambda_2t))
\end{split}
\end{align*}
Now solving the above system, we get
\begin{align*} 
\begin{split}
p^{\pm} = \frac{1}{\xi}{\mathcal{W}\big({\textstyle\frac{\xi}{4}}{\scriptstyle(\mathcal{R}_1^0-\mathcal{R}_2^0)}\big)} \pm {\frac{1}{4}}{(\mathcal{R}_1^0+\mathcal{R}_2^0)}{\exp\big[{-\mathcal{W}\big({\textstyle\frac{\xi}{4}}{\scriptstyle(\mathcal{R}_1^0-\mathcal{R}_2^0)}\big)}\big]}
\end{split}
\end{align*}
where $\mathcal{R}_1^0 = \mathcal{R}_1(u_1^0(x - \Lambda_1t),u_2^0(x - \Lambda_1t))$, $\mathcal{R}_2^0 = \mathcal{R}_2(u_1^0(x - \Lambda_2t),u_2^0(x - \Lambda_2t))$ and $\mathcal{W}$ is the Lambert W function \cite{corless1996lambertw}. We will plot the results of the non-tumbling case considering  a simple step function as the initial condition and compare the results with the corresponding full-particle simulations. We expect the left- and right-moving waves to be shifted with respect to the noninteracting case.   

\subsection{Biased random transitions: \texorpdfstring{$\lambda^{\pm} \neq 0$}{Lg}} 
When the system possesses an external signal, the particles begin their random turns due to the source term $\vec{g}(x,\vec{u})$ on the right-hand side of (\ref{conserv_law}); hence, following the same reasoning as in zero turnings, the system of ODEs (\ref{ODE}) returns 
\begin{align*}
V_{i1}\frac{du_1}{dt}+V_{i2}\frac{du_2}{dt} = V_{i1}g_1+V_{i2}g_2 \quad \text{for}\quad i = 1,2
\end{align*}   
We can write the above system in a more concise form with the approximation to $\sqrt{b}$; the solution propagates according to the differential equations
\begin{equation} \label{riemann}
\begin{split}
\frac{d\textbf{R}}{dt} = e^{\frac{\xi u_2}{2}}\textbf{G}(u_1,u_2) \quad\text{along the characteristics}\quad\\
\frac{d\textbf{x}}{dt} = \textbf{Q}(u_1,u_2) 
\end{split}
\end{equation}
where $\textbf{R} = (\mathcal{R}_1,\mathcal{R}_2)$ as defined in (\ref{R1R2}), $\textbf{G}$ is the source term whose entries are $V_i\cdot\vec{g}=\mu_1u_2 + \mu_2u_1$ for $i=1,2$ and $\textbf{Q}(u_1,u_2)=(\Lambda_1,\Lambda_2)$. The equations are integrable along the characteristics. So computing the solution of the kinetic model (\ref{kinetic_1}) is equivalent to numerically generating the characteristic paths in spacetime. \\
In the following section, we discuss the methods of solving the full-particle systems for both point and finite-size particles, followed by matching the time-dependent solutions with those of the numerical integration.

\subsection{Numerical examples}
\begin{figure*}
\centering
\includegraphics[width=\textwidth]{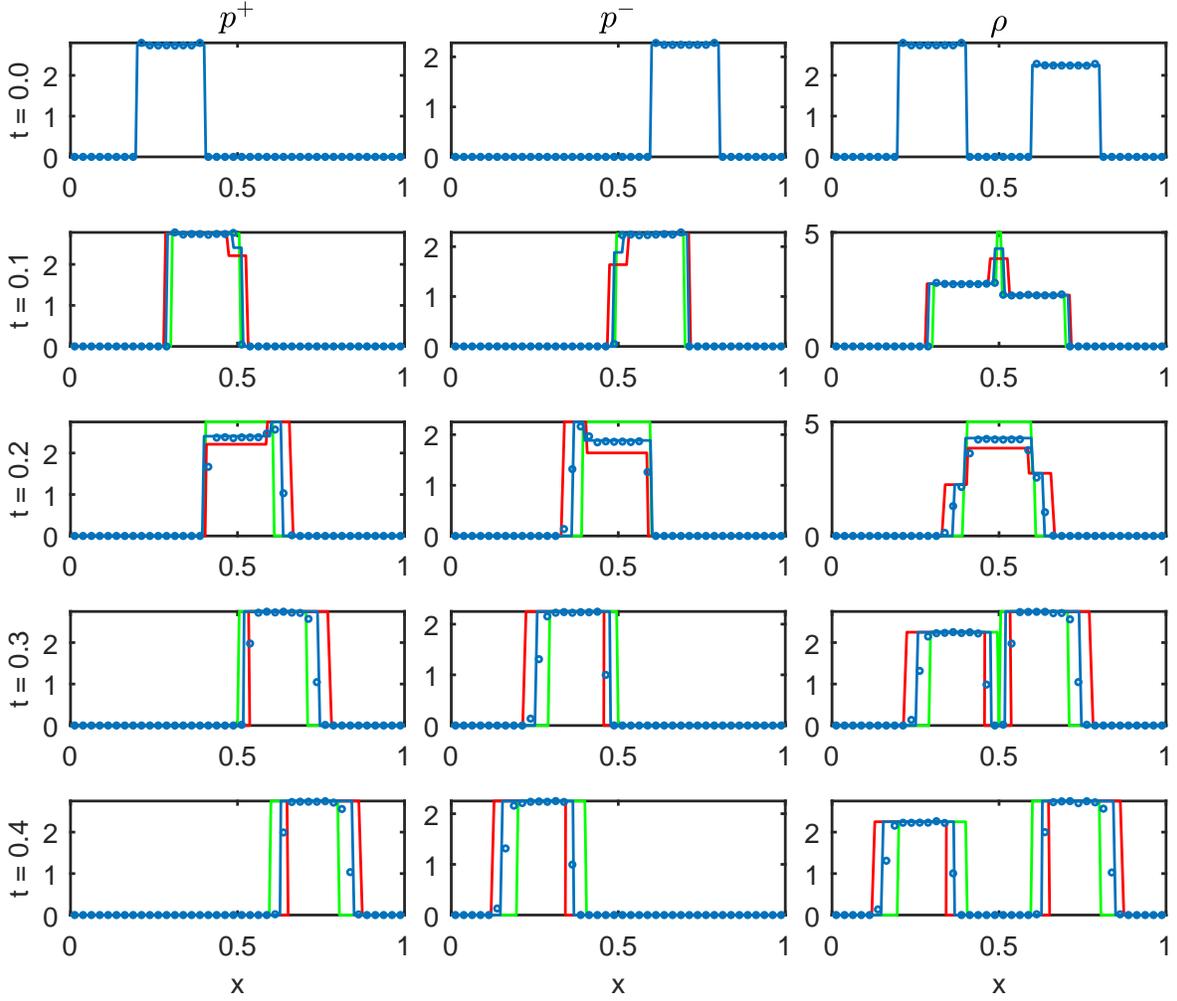}
\caption{Transient marginal densities of the kinetic model (\ref{kinetic_1}) for $\lambda^{\pm}=0$ when $\epsilon = 0$ (green line), $\delta(0.002,0.01) = $ (blue line) and $\delta(0.002,0.004) = 1$ (red line). Numerical results in the first and second columns show the band travels to the right and left, respectively, and the third column gives the total density $\rho = p^++p^-$. The particle simulations (circles) are obtained by $5000$ realisations. We use the initial condition (\ref{IC}), $N = 100$ and $c = 1$.}
\label{narrow_transient_Nosource}
\end{figure*}

\begin{figure*}
\centering
\includegraphics[width=\textwidth]{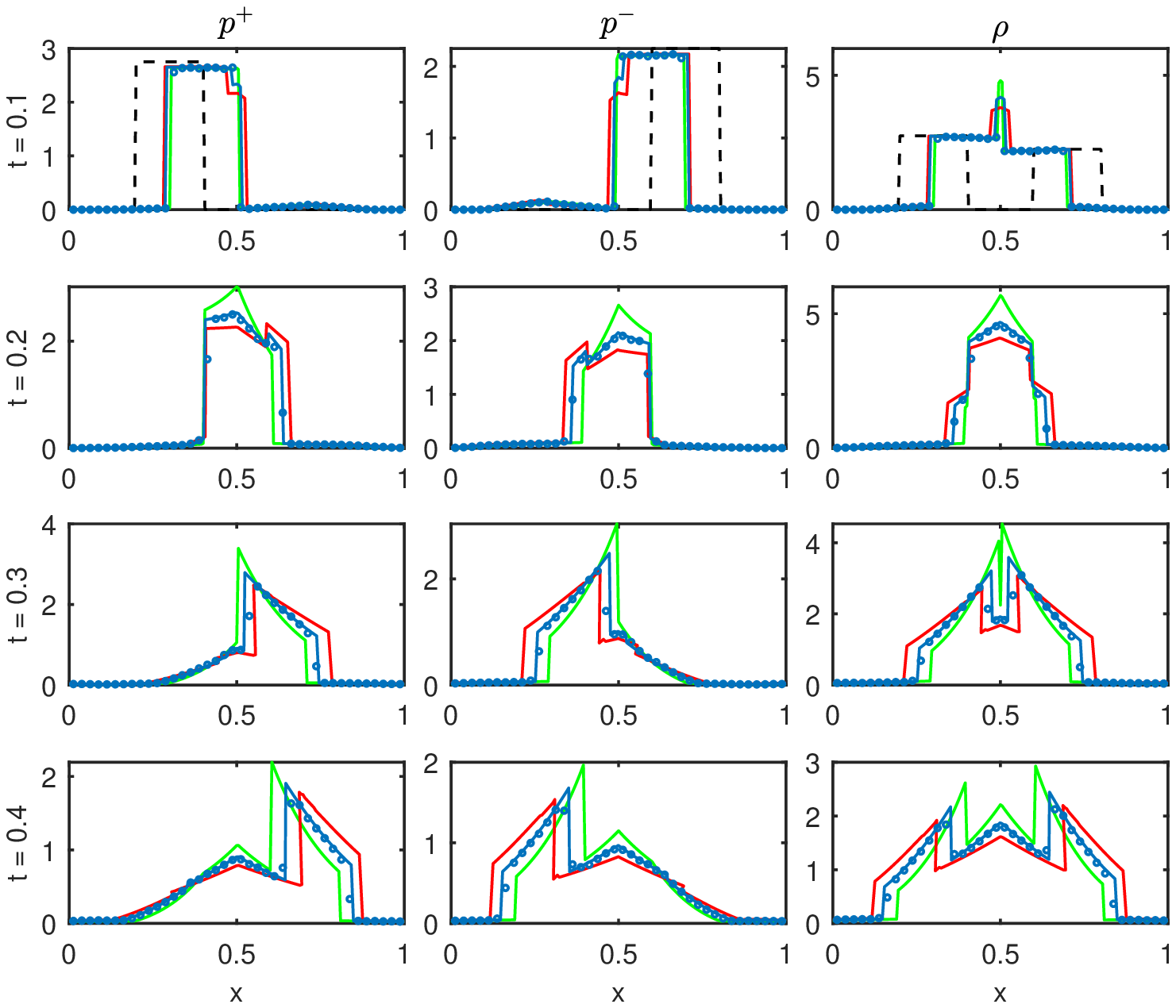}
\caption{Transient marginal densities of the kinetic model (\ref{kinetic_1}) for the signal function $S_1$ (\ref{signals}) when $\epsilon = 0$ (green line), $\delta(0.002,0.01) = $ (blue line) and $\delta(0.002,0.004) = 1$ (red line). Numerical results in the first and second columns show bands travelling to the right and left, respectively, and the third column gives the total density $\rho = p^++p^-$. The particle simulations (circles) are obtained by 5000 realisations. We use the initial condition (\ref{step_fun}), $N = 100$ and $c = 1$.}
\label{narrow_transient_source_S1}
\end{figure*}

\begin{figure*}
\centering
\includegraphics[width=\textwidth]{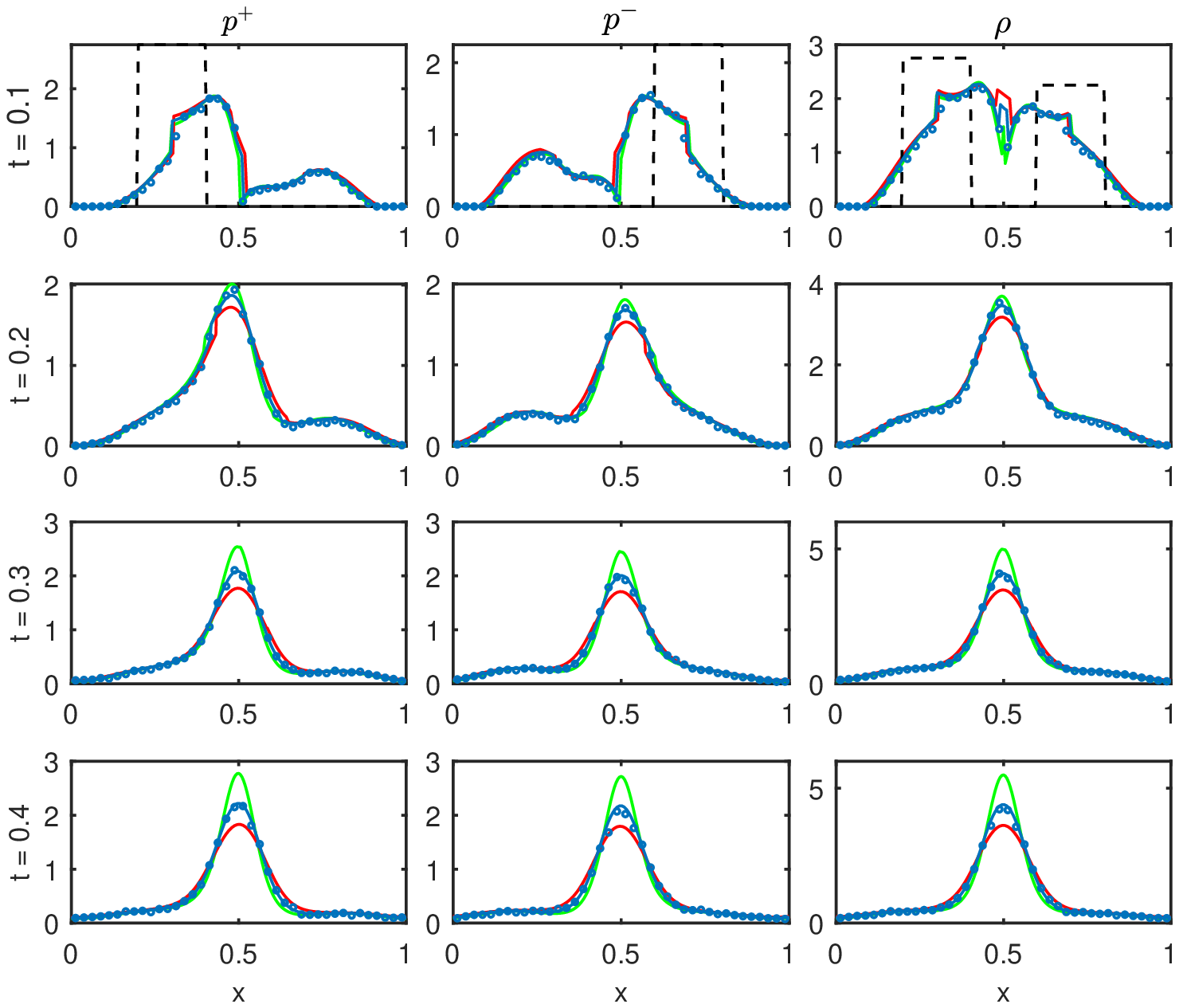} 
\caption{Transient marginal densities of the kinetic model (\ref{kinetic_1}) for the signal function $S_2$ (\ref{signals}) when $\epsilon = 0$ (green line), $\delta(0.002,0.01) = $ (blue line) and $\delta(0.002,0.004) = 1$ (red line). Numerical results in the first and second columns show the band travels to the right and left, respectively, and the third column gives the total density $\rho = p^++p^-$. The particle simulations (circles) are obtained by 500 realisations. We use the initial condition (\ref{step_fun}), $N = 100$ and $c = 1$.}
\label{narrow_transient_source_S2}
\end{figure*}

The task is now to assess the validity of the kinetic model from the transient solutions. The solution procedures studied under previous sections are illustrated through practical numerical examples, and the numerical results of the nonlinear kinetic model (\ref{kinetic_1}) are compared with particle simulations. However, since an analytical solution is not achievable with varying turning rates over $x$, we resort to a numerical integration method along the characteristics (see Appendix \ref{N_integration} for details). This way, we avoid problems that occur when using standard numerical methods for solving PDEs. We have already established the equations for this numerical integration in the previous section, where we have found a set of ODEs for the Riemann variables $\mathcal{R}_1$ and $\mathcal{R}_2$ on the characteristic directions. Given the initial conditions, we can obtain the solutions elsewhere by integrating along the characteristic curves, and the numerical procedure is the simple Euler's approximation. Each integration step requires information carried by both sets of characteristics, and they themselves depend on both $u_1$ and $u_2$ which leads to a nonuniform grid.\\
To generate point-particle simulations, we apply a simple time-stepping algorithm where a particle positioned at $X(t)$ evolves according to 
\begin{equation} \label{SSA}
X(t+\Delta t)=X(t)+V(t)\Delta t
\end{equation}
at a fixed time step $\Delta t$. The initial positions $X(0)$ are drawn from the initial density $P_0(\vec{x},\vec{v})$. A point particle which undergoes a velocity jump process has only two possible events; reflection near boundaries and random turns. The reflective boundary conditions can be implemented as follows:
\begin{equation*} \label{RBC}
\begin{aligned}
\text{If} \quad X(t+\Delta t)  < 0, \quad X(t+\Delta t) &= -X(t)-V(t)\Delta t; \\
\text{If} \quad X(t+\Delta t)  > L, \quad X(t+\Delta t) &= 2L-X(t)-V(t)\Delta t; \\
\text{complemented with}\quad V(t+\Delta t) &= -V(t).
\end{aligned}
\end{equation*}
The above conditions are used to avoid particles moving outside the domain walls when stepping forward in time. However, they do not generate new positions; instead, they detect the wall, switch the velocity and continue the random walk (that performed outside walls) in the opposite direction. The final step is the execution of the random velocity jumps. During the time interval $(t+\Delta t)$ a particle will turn with the turning probability $\lambda\Delta t$; otherwise, it advances with the same velocity. For finite-size particles we use an event-based algorithm that fits into the general class of  kinetic Monte Carlo (KMC) methods. It is an improved algorithm that counts all the interaction times in the system to produce time steps \cite{lubachevsky1991simulate}. As in a collision system, there are three events in the problem, except now there is a chance for particles to pass each other during an interaction. For this reason, a particle can bypass its neighbours and interact with a distant individual or with the walls. Therefore, it is necessary to check the interactions of each particle with every other particle and the walls. \\
The numerical examples presented in this section aim to do the following: illustrate the behaviour of the systems under biased and unbiased conditions, investigate whether the model and associated discrete processes support travelling wave solutions, and examine the effect of changing parameters, such as the size of the particles and collision probability, on the solutions. Unless explicitly stated otherwise, we consider a set of $N = 100$ particles with speed $c=1$ and chemotactic sensitivity coefficient $\chi=1$, placed inside a channel of length one ($L=1$). Given the size of the indistinguishable particles and the width of the channel, the formula in Appendix \ref{Collision_Prob} calculates the collision probability $\delta$. To bias the interactions, we consider two forms of signal functions: 
\begin{align} \label{signals}
\begin{split}
S_1(x)=1-2|x-0.5| \quad &\text{with} \quad\lambda_{0} = 2.5 \\
S_2(x)= 2e^{-50\,(x \,-\, 0.5)^2} \quad &\text{with} \quad\lambda_{0} = 20
\end{split}
\end{align} 
These signals are simple domains where attractants are occupied at a global maximum $0.5$. In this way, we develop insight into the impact of different signal gradients and baseline frequencies on the behaviour of a uniform concentration of particles. Besides, our results can be easily compared with those observed in the collision system. \\ 
We plot a subpopulation of right-moving $p^+$, a subpopulation of left-moving $p^-$ and the total population $\rho = p^++p^-$ for a simple non-tumbling case in Figure (\ref{narrow_transient_Nosource}), and with biases in Figures (\ref{narrow_transient_source_S1}) and (\ref{narrow_transient_source_S2}) up to time $t=0.4$. The initial particle arrangement is given by 
\begin{align} \label{step_fun}
p^+_0(x) = \frac{55}{100}\mathds{1}_{[0.2,0.4]} \quad\text{and}\quad p^-_0(x) = \frac{45}{100}\mathds{1}_{[0.6,0.8]}
\end{align}
where $\mathds{1}_A$ denotes the characteristic function of a set $A$. In the event-driven algorithm, the initial positions are generated randomly, avoiding overlaps; the collision probability will take care of the overlaps during subsequent iterations. The histograms are produced by dividing the domain into 40 bins. At each step, we check the number of particles in each bin; after that, the cumulative average is calculated, dividing the resulting value in each bin by the number of steps, total particle count and bin width. For the two signal functions $S_1$ and $S_2$, we performed $5000$ and $500$ realisations, respectively. Effectively, this implies that we are using $5\times 10^5$ ( and $5\times 10^4$) trajectories of all $N$ particles to compute the one-particle distribution histogram. \\
We compare the numerical predictions of the narrow channel system with those corresponding to both the point particles and the collision models. They are the reference points that allow us to see the competition between the most favourable signal environment and the volume exclusion of finite-size particles. In all figures, we follow the colour code: \textcolor{green}{green}, the point particle system ($\epsilon = 0$); \textcolor{blue}{blue}, the narrow channel system ($l > 2\epsilon$); \textcolor{red}{red}, the collision system ($l=2\epsilon$). The solid lines and the circles represent solutions for the PDE models and the KMC simulations, respectively. For more apparent graphs, we only include particle simulations of the narrow channel system. The particle size is fixed to $\epsilon=0.002$ so that $N\epsilon^2$ remains constant during model comparison. We then change the width of the channel to $0.004$ to form the collision system that gives $\delta(0.002,0.004) = 1$. This comparison explains the importance of excluded volume effects on the propagating wavefronts in biased and unbiased conditions.\\
The idea of distinct waves for subpopulations is pursued in \cite{erban2004individual} and \cite{simpson2013travelling}. The former referred to travelling bands of simple point particles, while the latter suggested those when the travelling wave speed coincides with the proliferative agent cell speed. The wavefronts of the point particle system (\ref{PP}) travel at constant speed $c$. Since $\epsilon=0$, overlap within the band is not a factor. However, when $\epsilon\neq 0 $, velocity changes due to interactions. From Fig. (\ref{narrow_transient_Nosource}) and Fig.(\ref{narrow_transient_source_S1}), we find the nonlinear system (\ref{kinetic_1}) obeys the noninteracting particles linear system up to $t=0.1$; thereafter, two fronts collide. Specifically, in the narrow channel, only $43.75\%$ collisions are involved in this. %The waves distort at $t=0.2$, recover the initial shape after bouncing off each other at $t=0.3$ and proceed in their respective directions ($t=0.4$). 
The rebounded waves at $t=0.3$ are shifted outward; apparently, this shift progresses with the increasing collision probabilities, which the eigenvalues, to $\mathcal{O}(\epsilon)$, $\Lambda^+ = c+c\xi p^- $ and $\Lambda^- = -c-c\xi p^+$ describe. When $\lambda\neq 0$, whether the particles are point or finite in size, they undergo instantaneous velocity changes; waves distort before collisions ( see Fig. (\ref{narrow_transient_source_S1}) and Fig.(\ref{narrow_transient_source_S2}) at $t=0.1$). The travelling bands are further disrupted due to interactions between finite-size particles, and the difference between the linear and nonlinear PDE solutions become more noticeable at $t=0.2$. \\
In general, we find higher densities around the peaks of the signal functions since both left and right moving individuals aggregate into favourable regions. However, the peaks are reduced for the narrow channel system and are further reduced for the collision system. This is because $56.25\%$ of overlaps raise the crowding in and around the centre of the narrow channel domain compared to the single-file channel. In other words, the higher the collision probability the lesser the peak. Moreover, the theoretical predictions for finite-size particles in the narrow channel compare well with their simulation counterparts. This is mainly because of the low occupancy of particles ($3.14\%$) in the domain with more overlaps. We required significantly fewer simulations for the signal function $S_2$ compared to $S_1$, as particles reorganized themselves rapidly under higher baseline frequency. For the same reason, we do not observe continuing kinetic waves in Fig.(\ref{narrow_transient_source_S2}).

\section{Stationary solution} \label{sec_stat}
The stationary solution of (\ref{kinetic_1}) with no-flux boundary conditions can be obtained by solving the following ODE for $p_{st}$: 
\begin{equation*} \label{20}
c\frac{dp_{st}}{dx}+c\xi p_{st}\frac{dp_{st}}{d x}+(\lambda^+-\lambda^-)p_{st} = 0
\end{equation*} 
The $\epsilon=0$ returns the solution $p_{st} = A\exp\left[{-\int^x_0\frac{\Uplambda(u)}{c}du}\right]$ for point particles, while for finite-size particles this would be $p_{st} = \frac{1}{\xi}\mathcal{W}\left(A\xi\exp\left[{-\int^x_0\frac{\Uplambda(u)}{c}du}\right]\right)$, where $\Uplambda = \lambda^+-\lambda^-$, $\mathcal{W}$ is the Lambert W function and $A$ is a constant to be determined using the normalisation condition.\\
The stationary solution $P_{st}$ for the full particle system derives from the fact that in two dimensions (two particles), the inner region is a diagonal of width $2\epsilon$, and the inflow towards the excluded region changes from factor $(1-\delta)$. In three dimensions (three particles), a slab of the same width along the diagonals represents the inner regions.
\begin{figure}
\centering
\includegraphics[scale=0.8]{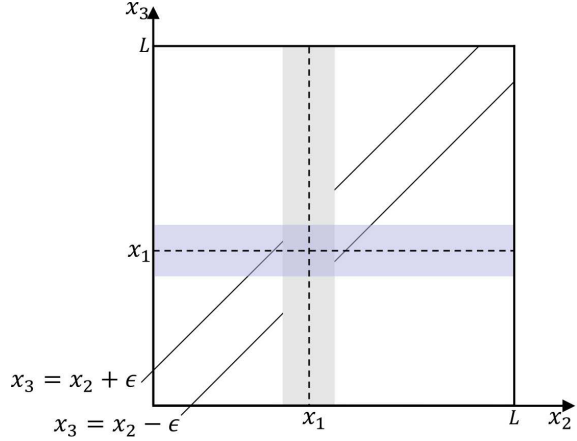} 
\caption{Schematic of the excluded domain for three interacting particles when one particle is fixed at $x_1$. The shaded areas show the inner slabs of width $2\epsilon$.}
\label{Configuration3particles}
\end{figure}
As depicted in Figure (\ref{Configuration3particles}) when one particle is fixed at $x_1$, the inflow changes from $1-\delta$ when first enters from the inner slab of the second particle and then from $(1-\delta)^2$ further from the inner slab of the third particle. If the system has $n (<N)$ interacting particles, one particle has $n-1$ inner regions.\\
Thus, for position index $i\neq j$, define 
$$
\phi(x_i,x_j) = \left\{
   \begin{array}{cc}
     1 & |x_i-x_j|\leq \epsilon \\
     0 & |x_i-x_j|> \epsilon   
   \end{array}
   \right. 
$$
Then $$n(\phi)=\sum\limits^{N-1}_{i=1}\sum\limits^N_{j=i+1} \phi(x_i,x_j),$$
and we can write the stationary solution as 
\begin{align} \label{SS1}
P_{st}(\vec{x}) = A[1-\delta]^{n(\phi)}\exp\left[{-\sum \limits_{i=1}^N \int \limits_0^{x_i}\frac{\Uplambda(u)}{c}du}\right]
\end{align}
Note that, the collision probability $\delta$ is invariant to switching of particles, and we still adhere to the low volume fraction assumption. When $\delta\rightarrow 0$, we recover the steady state for noninteracting particles. Integrating (\ref{SS1}) in higher-dimension and direct calculation of the arbitrary constant $A$ may not be feasible; instead, we move to the application of the Metropolis-Hastings (MH) algorithm that allows us to sample directly from the $N$-dimensional microscopic density \cite{metropolis1953equation}. Further details are given in Appendix \ref{MHA}.\\
The Fig.(\ref{narrow_stationary}) shows the model and simulation results for both point and finite-size particles with unit speed.
\begin{figure}
\centering
\includegraphics[scale=0.9]{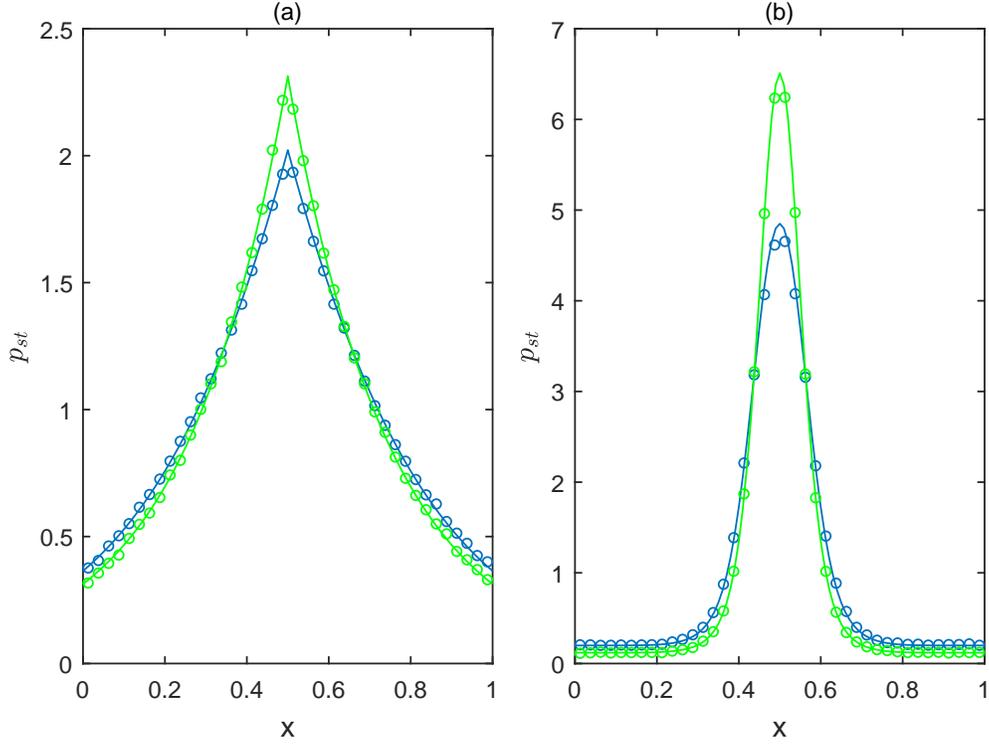}
\caption{Stationary marginal density $p_{st}$ (solid lines) and the simulations from the MH algorithm (circles) for $\epsilon=0$ (green) and $\epsilon=0.002$ (blue). (a) Solutions corresponding to the signal function $S_1$. (b) Solutions corresponding to $S_2$.}
\label{narrow_stationary}
\end{figure}
With $N = 100$ particles of size $\epsilon = 0.002$ in a narrow channel of length $L=1$, this corresponds to $0.000314/l$ fraction of filled volume. The width of the channel determines the collision probability; for $l>0.004$, particles can pass each other, while $l=0.004$ turns it into a full-collision system. In accordance with the transient solutions, we choose $l = 0.01$ which returns $\delta(0.002,0.01) = 0.4375$. The histograms for the MH algorithm are produced by dividing the domain into 40 bins, and initially the particles are evenly spaced in the domain. An acceptance rate in the $0.1$ order of magnitude, and $10^6$ steps of the algorithm produce the desired results. We monitor the number of particles in each bin at every step: when a proposed move is rejected, the old configuration is added over to the count, whereas if the move is accepted, the new configuration is added. At the end of the process, the cumulative average is calculated; dividing the resulting value in each bin by the number of steps, total particle count and bin width.\\
The stationary solution of the kinetic model agrees well with the particle simulation results for both interacting and noninteracting systems. As in time-dependent solutions, we see a lower density around the peak of the signal functions for finite-size particles. It is clear from the figure that the narrow channel allows some overlaps based on the passing probability around the maximum point of the signal while the colliding individuals redistribute to other accessible areas in the domain. Even though we did not include stationary solutions from the collision system, we expect a much lower peak in this situation unless for considerably smaller particle size.

\section{Conclusion}
This paper has considered a system of $N$ identical hard-cores of size $\epsilon$ with fix speed $c$ that undergoes a velocity-jump process in a narrow bounded channel. These random changes in the velocity are instantaneous and distributed according to a Poisson process. The finite-size of particles means that the motion is correlated; the interactions in the channel give rise to the so-called interface conditions. From a high-dimensional PDE system, we have obtained the kinetic model under the small volume fraction assumption whilst considering the interactions at the particle level. The approach was simply based on a regular perturbation problem. The derived equations are nonlinear in the transport term, and incorporate a collision probability that resembles a collision system at its maximum value. We have verified the model with numerical simulations, comparing its solutions with the corresponding stochastic simulations of the underlying particle system as well as against the interaction-free linear system. The plots confirm that the model captures the features at the particle level well. Besides, we have implemented both time-dependent and stationary simulations of the system. The time-dependent solutions are non-dissipative as we have considered a systematic approach based on characteristics for hyperbolic balance laws. Note that, we have not commented on the collision system's outputs as they are examined already in \cite{ralph2020one} following different conditions. In fact, one can refer to this study for more details regarding the asymptotic analysis and simulation algorithms. Although our method developed here is in its simplest setting, this core model constitutes the first step towards extensions in many directions. We will be considering distinguishable particles in a narrow channel in our forthcoming paper. Another interesting but complicated extension is the consideration of anisotropic particles to examine how the transport model changes with noncircular particles \cite{peng2016diffusion}.   

\begin{acknowledgments}
We would like to thank Tertius Ralph for several helpful discussions. 
\end{acknowledgments}

\appendix
\section{Collision Probability} \label{Collision_Prob}
In general, the collision probability depends on the sizes of the species and the container. Here we consider a single species of diameter $\epsilon $ inside a narrow channel of width $l$, such that $2\epsilon \leq l$. When one particles is at $y$, where $Y\sim U(\frac{\epsilon}{2},l-\frac{\epsilon}{2})$, 
\begin{figure}
\centering
\includegraphics[scale=0.6]{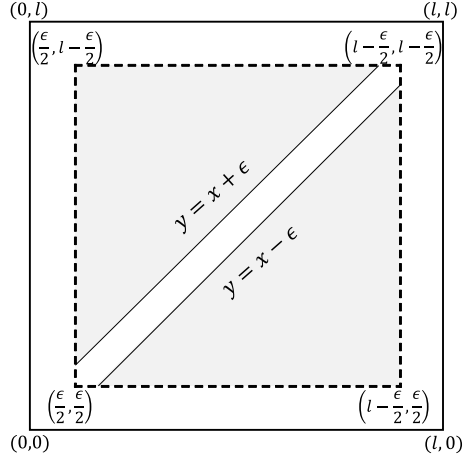} \caption{The collision square illustrates the space available for the red particle when the blue particle is fixed at $y$. The upper left and the lower right triangles represent, respectively, passing below and passing above.}
\label{Collision square}
\end{figure}
the space available for the second particle to unobstructedly move is given by the top and the bottom triangles of Fig.(\ref{Collision square}). In other words, the probability of particle $2$ passing particle $1$; below (upper left triangle) is $\frac{max\left\{0, y-\frac{3\epsilon}{2}\right\}}{l-\epsilon}$ and above (lower right triangle) is $\frac{max\left\{0, l-y-\frac{3\epsilon}{2}\right\}}{l-\epsilon}$. The position density function of particle 1 is the constant function $1/(l-\epsilon)$. Hence, the total probability of particle 2 passing particle 1 is   
\begin{align*}
\int\limits^{(l-\frac{\epsilon}{2})}_{\frac{3\epsilon}{2}} P(\text{passing below}|Y=y)f_Y(y)dy &+\\ \int\limits^{(l-\frac{3\epsilon}{2})}_{\frac{\epsilon}{2}} P(\text{passing above}|Y=y)f_Y(y)dy &= \frac{(l-2\epsilon)^2}{(l-\epsilon)^2}
\end{align*}
Since the collision probability is ($1-$ the passing probability), we remark the following special case:
\begin{enumerate}
\item When $\epsilon \rightarrow 0$, collision probability vanishes; hence, we obtain an noninteracting system,
\item When $l = 2\epsilon$, particles cannot pass each other; hence, the system becomes a collision system.
\end{enumerate}

\section{Numerical Integration} \label{N_integration}
Here we detail the numerical integration procedure along the characteristics according to the system (\ref{riemann}).
\begin{figure}
\centering
\includegraphics[scale=0.6]{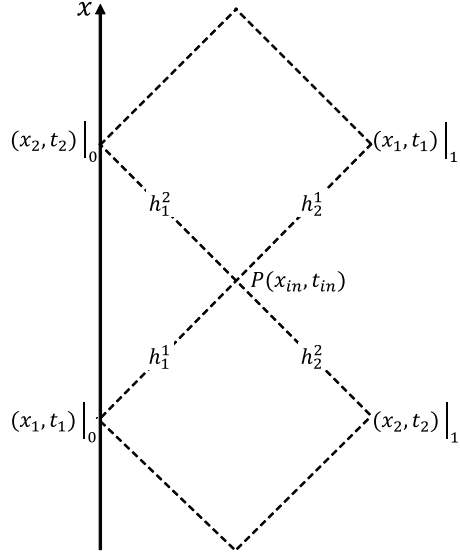}
\caption{Two characteristics originating from $\textbf{x}_0=(x_1,x_2)|_0$ at time $\textbf{t}_0=(t_1,t_2)|_0$ intersect at $P(x_{in},t_{in})$, then reach $\textbf{x}_1=(x_1,x_2)|_1$ at $\textbf{t}_1=(t_1,t_2)|_1$.}
\label{character1}
\end{figure} 

\begin{enumerate} \label{EU_alg}
\item[S1] Determine the time and position of the intermediate crossover point $P(x_{in},t_{in})$ of the two characteristics originating from $\textbf{x}_0\equiv \textbf{x}(\textbf{t}_0)$ (see Fig. (\ref{character1})). Here we use the two equations; $x_{in}=\textbf{x}_0 + (h_1^1,h_1^2)\circ\textbf{Q}$, the output from the Euler's step and $t_{in} = \textbf{t}_0+(h_1^1,h_1^2)$, where $h_1^i$s are the time steps for their respective characteristics.
\item[S2] Calculate the Riemann variables $\textbf{R}$ at $P$ using
\begin{align*} 
\textbf{R}(x_{in},t_{in})=\textbf{R}(\textbf{x}_0,\textbf{t}_0) + (h_1^1,h_1^2)\circ\textbf{G}
\end{align*}    
Each element in $\textbf{R}$ is a combination of $u_1$ and $u_2$ along the two curves, which gives us two equations to solve and find the updated $u_1$ and $u_2$ at the crossover point.
\item[S3] Using the updated $u_1$ and $u_2$ calculate the new positions $\textbf{x}_1\equiv\textbf{x}(\textbf{t}_1)$ beyond the intersection point. Similar to equations in S$1$, again from the Euler's step we have  $\textbf{x}_1={x}_{in} + (h_2^1,h_2^2)\circ\textbf{Q}$, and $\textbf{t}_1 = {t}_{in}+(h_2^1,h_2^2)$.
\item[S4] Find $\textbf{R}$ beyond intersections using
\begin{align*} 
\textbf{R}(\textbf{x}_1,\textbf{t}_1) = \textbf{R}(x_{in},t_{in}) + (h_2^1,h_2^2)\circ\textbf{G}
\end{align*}    
\end{enumerate}
Here $h_1^i$ and $h_2^i$, for $i=1,2$ are calculated from the equations given in S1 and S3. The total time elapsed, say $h$, is the sum of $h_1^i$ and $h_2^i$ along each characteristic. We cannot maintain a fixed $h$ since $h_1^i$ and $h_2^i$ are constantly changing during the process. We may start off with a uniform grid at $t=0$, but subsequently follows the characteristics. Hence the curves will be approximate straight lines. 

\section{The Metropolis-Hastings algorithm} \label{MHA}
When $\Phi(\vec{x})=\sum \limits_{i=1}^N \int \limits_0^{x_i}\frac{\Uplambda(u)}{c}du$ be the energy associated with the configuration $\vec{x}\in\Omega^N$, the stationary density (\ref{SS1}) becomes 
\begin{equation} \label{P_st}
P_{st}(\vec{x}) = A[1-\delta]^{n(\phi)}e^{(-\Phi(\vec{x}))}\quad\text{for}\quad\vec{x}\in\Omega^N
\end{equation}
Note that $\Phi$ is not defined outside the domain; therefore, we set $\Phi(\vec{x})=\infty$ for $\vec{x}\notin\Omega^N$.
The MH algorithm samples configurations according to the density $P_{st}$ as follows:
\begin{enumerate}\label{MH-alg1}
\item[S1] Select a particle $i$ at random and calculate the close encounters with the other $\textbf{x}_j$s for $i\neq j=1,2,...,N$.
\item[S2] Generate a candidate $\textbf{y}_i=\textbf{x}_i+hX$ where $X\sim N(0,1)$ and $h$ a tunable parameter. 
\item[S3] Count the close encounters with $\textbf{y}_i$ and $\textbf{x}_j$s for $i\neq j=1,2,...,N$. (the difference of above counts in steps 1 and 3 $=(n-1)$)
\item[S4] Compute the difference $\Delta\Phi$ between the current and modified configurations
\item[S5] Accept $\textbf{y}_i$ with probability $p$ = min(1,$(1-\delta)^{n-1}$exp(-$\Delta\Phi$)) and set $\textbf{x}_{i+1}=\textbf{y}_i$, otherwise set $\textbf{x}_{i+1}=\textbf{x}_i$
\end{enumerate}
In steps S1-S3, we compare the close encounters before and after modifying each selection. In this way, we are counting two-particle overlaps.
%\pagebreak
%\bibliographystyle{plain}
%\bibliography{Ref}
\providecommand{\noopsort}[1]{}\providecommand{\singleletter}[1]{#1}%

\end{document}